# High-throughput development of flexible amorphous materials showing large anomalous Nernst effect via automatic annealing and thermoelectric imaging

Sang J. Park[1,*], Ravi Gautam[1], Abdulkareem Alasli[2], Takamasa Hirai[1], Fuyuki Ando[1], Hosei Nagano[2], Hossein Sepehri-Amin[1] and Ken-ichi Uchida[1,3,*]

[1] National Institute for Materials Science, Tsukuba 305-0047, Japan

[2] Department of Mechanical Systems Engineering, Nagoya University, Nagoya 464-8601, Japan

[3] Department of Advanced Materials Science, Graduate School of Frontier Sciences, The University of Tokyo, Kashiwa 277-8561, Japan

*Corresponding authors: PARK.SangJun@nims.go.jp (S.J.P.); UCHIDA.Kenichi@nims.go.jp (K.U.)

**Abstract**

This work demonstrates high-throughput screening of flexible magnetic materials for efficient transverse thermoelectric conversion based on the anomalous Nernst effect (ANE). The approach integrates automated annealing and contactless measurement of transport properties using lock-in thermography. We screen 151 Fe-based alloy ribbons with varying compositions and annealing conditions. Seven high-performance candidates with mechanical flexibility are identified, exhibiting anomalous Nernst coefficients of up to 4.8 µV/K, the highest value reported for flexible materials. Structural analysis reveals that ANE enhancement occurs universally near the first crystallization temperature of the Fe-based ribbons, without strong correlation with composition. Notably, the enhancement is also observed in samples without Cu or Fe nanoclusters, indicating that short-range atomic order in the amorphous matrix may play a role in ANE. These findings demonstrate the effectiveness of high-throughput methodologies for discovering advanced ANE materials and provide new insights into thermoelectric conversion in disordered systems where conventional design principles fall short.

**Introduction**

Global warming has accelerated the transition toward a carbon-neutral society, emphasizing the urgent need for sustainable energy solutions. A significant portion of energy is dissipated as waste heat during energy conversion processes, representing a major loss of potentially usable energy. Consequently, harvesting energy from low-grade waste heat has emerged as a critical strategy for achieving a sustainable future. In this context, thermoelectrics have been recognized as a promising solution for converting waste heat into usable electrical energy, offering advantages such as vibration-free operation and the absence of environmentally harmful cooling fluids [1–4].

Recently, transverse thermoelectrics based on the anomalous Nernst effect (ANE) have gained significant attention [5–10] due to their potential to address the challenges faced by conventional Seebeck-effect-based longitudinal thermoelectric devices [1–4,11–13]. In the ANE, the electric charge is generated perpendicular to the applied heat flux, offering various benefits in device configuration. One key advantage is that transverse thermoelectric devices do not require junctions or additional electrode materials, such as solder, for their construction. As a result, the devices can be fabricated from a single type of material (either n- or p-type) [14–18], thereby eliminating performance degradation due to contact resistances and the need to match the mechanical properties of different materials (e.g., thermal expansion coefficients) [11,19]. Moreover, the output voltage and power scale with the transverse length and area, respectively. These characteristics enhance device scalability, making such devices more suitable for large-scale energy harvesting applications, despite their relatively low conversion efficiency compared to longitudinal counterparts [7–10].

Over the last decade, significant efforts have been made to enhance the thermoelectric conversion performance of ANE, focusing specifically on improving the transverse transport mechanisms of electrons within solids. As proposed in the anomalous Hall effect (AHE), the mechanisms can be classified into intrinsic and extrinsic components [20–24]. Previous studies have primarily focused on discovering new materials with large intrinsic contributions, described by the Berry curvature, a geometrical property that characterizes the energy eigenstates across the electronic band structure [25–27]. Topological materials, such as Weyl semimetals, have exhibited large ANE due to their unique electronic band topology. On the other hand, evidence for extrinsic contributions to ANE has been reported in a few studies [6,28,29].

Despite the intensive efforts, significant challenges remain for practical implementation. According to the Mott relation, the intrinsic ANE is governed by the Berry curvature averaged over the states with finite entropy driven by thermal fluctuations [27,30,31]. This implies that the intrinsic contribution is highly sensitive to the position of the electronic band relative to the Fermi level. As a result, intrinsic ANE typically requires high-quality, clean samples with precise tuning of chemical composition to control the Fermi level. Additionally, many single-crystalline materials exhibiting large intrinsic ANE possess structural anisotropy [6,28,32], necessitating well-aligned crystal orientations relative to the applied heat flux and output voltage, and often suffer from limited mechanical robustness or flexibility. While some topological polycrystals have demonstrated ANE values comparable to those of single crystals, these cases are limited to isotropic structures, such as cubic systems [33–35]. For extrinsic ANE, the reported signal intensity at room temperature is relatively low compared to the intrinsic counterpart [6,28,29] and is known to dominate in super-clean materials (i.e., those with extremely low electrical resistivity [36]), thereby facing limitations similar to those of intrinsic ANE in terms of scalability and practical applicability.

To realize ANE-based energy harvesting applications, new breakthroughs in material design are essential, moving beyond the conventional approaches that rely on high-quality single-crystalline materials with limited compositional and structural diversity. Recently, significant ANE enhancements have been reported in fully amorphous or partially crystallized materials within an amorphous matrix, which offer superior mechanical flexibility and device compatibility [37–41]. For instance, Ravi et al. developed flexible ANE materials with an anomalous Nernst coefficient ($S_{ANE}$) of 3.7 µV/K in annealed amorphous alloys, having the composition $Fe_{84.7}Si_{2.8}P_{3.8}B_{7.8}Cu_{0.7}C_{0.2}$ (at%), approximately 70% higher than that of their fully amorphous counterparts [37]. While this enhancement was phenomenologically explained by the presence of precipitated Cu nanoclusters, similar improvements, with $S_{ANE}$ of 3.7 µV/K, were consistently observed in amorphous alloys without Cu, composed of $Fe_{92.5}Si_5B_{2.5}$, and attributed instead to the formation of amorphous–crystalline hetero-domain that generates transverse deflection [38]. This mechanism was supported by a theoretical model, which suggested that transverse deflection may arise through a distinct mechanism from conventional intrinsic and extrinsic mechanisms when two domains with distinct transport properties are physically mixed. While the detailed mechanisms in such complex disordered systems remain to be fully understood, this approach holds promise by enabling the use of non-single-crystalline materials in developing efficient ANE materials, thereby overcoming key limitations of conventional approaches. However, to make this technology viable, further improvements are necessary to reach performance levels comparable to or exceeding those of conventional single-crystal-based approaches. Achieving such enhancements requires an in-depth understanding of how the amorphous–crystalline transition affects transport properties, particularly within complex composite systems.

Our approach to accelerating the development of efficient non-single-crystalline ANE materials is the high-throughput screening of transverse thermoelectric materials. This high-throughput screening is based on the integration of annealing condition optimization using an automatic quenching system and contactless transverse thermoelectric property measurements using lock-in thermography (LIT, **Fig. 1**). The automated system enables precise sample annealing without human labor or error (**Fig. 1a**). LIT enables simultaneous, contact-free measurements of transport properties, eliminating the need for individual heater or electrode attachment for each sample (**Fig. 1b**). In this study, using this high-throughput screening method, we evaluated 151 samples with varying compositions and annealing conditions, and identified 7 promising candidates representing approximately the top 5% in transverse thermoelectric performance. Notably, all of the promising samples retained mechanical flexibility due to their mild annealing conditions, enabling the versatile applications to various heat sources, such as curved heat surfaces [42]. Detailed analyses of these samples revealed $S_{ANE}$ values reaching 4.8 µV/K, a state-of-the-art value among flexible materials reported in both bulk and thin-film form.

Furthermore, statistical analysis of the large dataset revealed that ANE enhancement universally occurs near the crystallization temperature. Atom probe tomography (APT) and scanning transmission electron microscopy (STEM) confirmed that such enhancement can arise even without distinct crystalline order or the formation of nanoclusters, highlighting the importance of nanostructure engineering as well as the control of possible short-range ordering in transverse thermoelectric conversion within disordered systems. Our high-throughput approach and results mark a new phase in the development of highly efficient ANE materials based on non-single-crystalline platforms and hold strong potential to advance materials discovery, for example via data-driven machine learning, and device optimization for transverse thermoelectric applications.

# Results and discussion

## Strategy for high-throughput screening of transverse thermoelectric materials

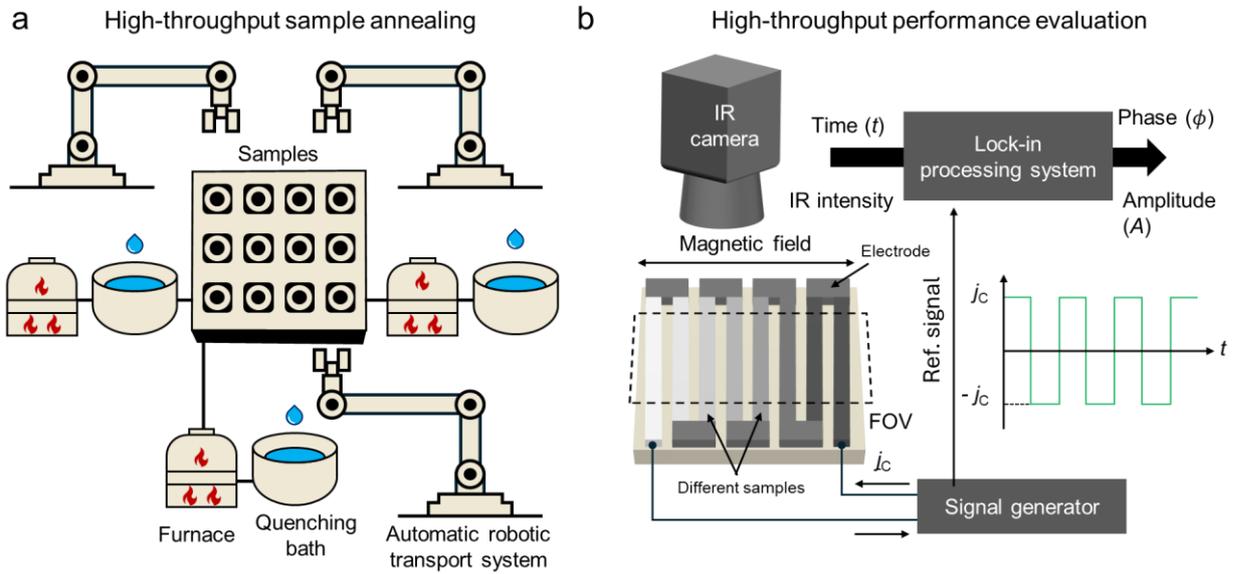

**Fig. 1 | Schematic illustration of high-throughput screening of transverse thermoelectric materials**. (a) An automatic annealing system for high-throughput sample processing. (b) lock-in thermography for high-throughput evaluation of magneto-thermoelectric performance.

We begin by outlining the high-throughput screening strategy for discovering efficient transverse thermoelectric materials. Traditionally, the development of bulk polycrystalline ANE materials involves a feedback-based fabrication process with several key steps [33,35,37–39]: (1) pelletization of constituent elements to achieve the desired compositions and phases, including pressing, alloying, melting and sintering; (2) post heat treatment, if necessary, including annealing to enhance crystallinity; (3) evaluation of ANE properties by measuring the voltage under a temperature gradient ($\nabla T$) and magnetic field ($H$); and (4) revisiting step (1) or (2) to optimize fabrication conditions or tune chemical composition. For annealed amorphous alloys, where ANE performance strongly depends on the annealing temperature ($T_a$)-controlled structural changes and peaks at a moderate $T_a$ (e.g., 653–698 K in refs. [37,38]), traditional methods face significant bottlenecks in steps (2) and (3), especially due to the large number of samples ($n$) with varying $T_a$. For example, post-annealing [step (2)] is typically conducted in a furnace, which requires manual operation, limiting high-throughput sample fabrication and potentially introducing human error. Additionally, vacuum sealing is often required for samples sensitive to oxidation. Moreover, the throughput of the ANE measurement and evaluation [step (3)] is constrained by conventional electrode-contact-based methods, where the electrical voltage is measured under $\nabla T$ and $H$. To measure the ANE, at least two electrodes and one heater must be attached to each sample in the transverse and longitudinal directions, respectively, with reproducible electrical and thermal contact conditions. This setup restricts ANE performance evaluations to a single sample at a time, requiring significant human intervention and greatly limiting throughput efficiency. Consequently, conventional ANE studies have often reported values based on only a few samples with selected optimization parameters [33,35,37–39], potentially overlooking the highest achievable performance of the material.

We overcome the bottlenecks in steps (2) and (3) by employing the following strategies. For post-annealing [step (2)], we utilized a customized automatic annealing system that performs heat treatments automatically (**Fig. 1a**). The system consists of three sets of robotic arms, quenching baths, and furnaces, and operates independently of human intervention, enabling high-throughput investigation of the annealing temperature dependence. The robotic arms are connected to a high-vacuum pump to protect oxidation-sensitive samples during annealing. The detailed operation procedure, including heat treatment, gas exchange, and water quenching, is described in **Methods**. All operations are controlled by a programmable centralized unit. The system can load up to 12 samples simultaneously and anneal three samples at once.

In addition, we overcome the bottleneck in evaluating transverse thermoelectric properties [step (3)] by measuring the anomalous Ettingshausen effect (AEE), a transverse charge-to-heat conversion in magnetic materials. The AEE is directly proportional to the ANE through Onsager reciprocity, expressed by the following relationship:

$$S_{\mathrm{ANE}} = \Pi_{\mathrm{AEE}}/T, \qquad (1)$$

where $\Pi_{\mathrm{AEE}}$ and $T$ represent the anomalous Ettingshausen coefficient and absolute temperature, respectively. This reciprocal relation allows the high-throughput screening of materials exhibiting large ANE by measuring AEE. To measure AEE, we employed LIT, which captures the temperature difference $\Delta T$ generated in the transverse direction by applying a longitudinal charge current $j_{\mathrm{C}}$ (**Fig. 1b**). LIT enables high-throughput evaluation by simultaneously collecting data from multiple points within a field-of-view (FOV, **Fig. 1b**), as previously demonstrated in composition-spread films [43–47]. The $\Delta T$ is detected using an infrared (IR) camera, synchronized with an AC signal generator and lock-in processing system, achieving a high temperature resolution (< 1 mK) [48]. This approach eliminates the need to attach a heater and electrodes to each sample individually. The FOV covers an area of 7.68 mm × 9.60 mm with 327,680 data points (512 × 640 pixels), allowing simultaneous detection of transverse thermoelectric signals from multiple samples connected electrically in series (**Fig. 1b**).

We measured the AEE signal using LIT in the following way. A square-modulated AC charge current is applied to the samples with applying steady $H$. The IR camera captures the lock-in amplitude ($A$) across the transverse direction (sample thickness direction in **Fig. 1b**), and the phase ($\phi$) of temperature modulation relative to the applied AC current. The measured signal includes contributions from the AEE ($\propto j_C$), the Peltier effect ($\propto j_C$), and Joule heating ($\propto j_C^2$). The Joule heating contribution is separated by applying a zero-offset square-wave-modulated current (**Fig. 1b**), producing a constant background signal ($\propto j_C^2$) without periodic temperature variations. To isolate the pure AEE contribution, the signals are symmetrized with respect to the magnetic field ($\pm H$), resulting in the field-odd components $A_{\text{odd}}$ and $\phi_{\text{odd}}$, defined by the following equations [45,49–53]:

$$A_{\text{odd}} = \frac{|A_{+H}\exp(-i\phi_{+H}) - A_{-H}\exp(-i\phi_{-H})|}{2}, \quad (2\text{-}1)$$

$$\phi_{\text{odd}} = -\arg\left[\frac{(A_{+H}\exp(-i\phi_{+H}) - A_{-H}\exp(-i\phi_{-H}))}{2}\right], \quad (2\text{-}2)$$

where $A_{+H(-H)}$ and $\phi_{+H(-H)}$ represent the $A$ and $\phi$ values at positive (negative) $H$, respectively. Note that the Peltier effect is independent of $H$ and the magneto-Peltier effect is symmetric with respect to $H$. Thus, their contributions do not appear in $A_{\text{odd}}$ and $\phi_{\text{odd}}$, enabling the pure detection of the AEE contribution. $\Pi_{\text{AEE}}$ is given by [49,50,52,53]:

$$\Pi_{\text{AEE}} = \frac{\pi}{4}\frac{\kappa \Delta T_{\text{AEE}}}{j_c t}, \quad (3)$$

where $\Delta T_{\text{AEE}}$ is the magnitude of $\Delta T$ induced by AEE along the thickness ($t$) direction (i.e., $\Delta T_{\text{AEE}} = 2A_{\text{odd}}$), and $\kappa$ is the thermal conductivity of the sample. Although most parameters in Eq. (3) can be evaluated at high throughput, measuring $\kappa$ across a large number of samples remains challenging. Therefore, we measured $\kappa$ and determined $\Pi_{\text{AEE}}$ values only for promising samples after the high-throughput screening based on the AEE measurements. We thus evaluate materials based on their charge-to-heat conversion performance ($\Lambda_{\text{AEE}} \equiv \frac{\nabla T_{\text{AEE}}}{j_c}$), defined as the transverse temperature gradient $\nabla T_{\text{AEE}} (= \Delta T_{\text{AEE}}/t)$ per input charge current density ($j_C$), without directly including $\kappa$. Detailed results for these selected samples are discussed in **Section "Quantitative evaluation of transverse thermoelectric performance in selected samples"**. It is also worth noting that the parameter $\Lambda_{\text{AEE}}$ directly represents the heat flux sensitivity of the material ($\propto \frac{S_{\text{ANE}}}{\kappa} = \frac{\pi}{4} \frac{\nabla T_{\text{AEE}}}{j_c} T \propto \Lambda_{\text{AEE}}$), which is another promising application area for ANE [9,16,54].

**Development of efficient ANE materials through feedback-based high-throughput process**

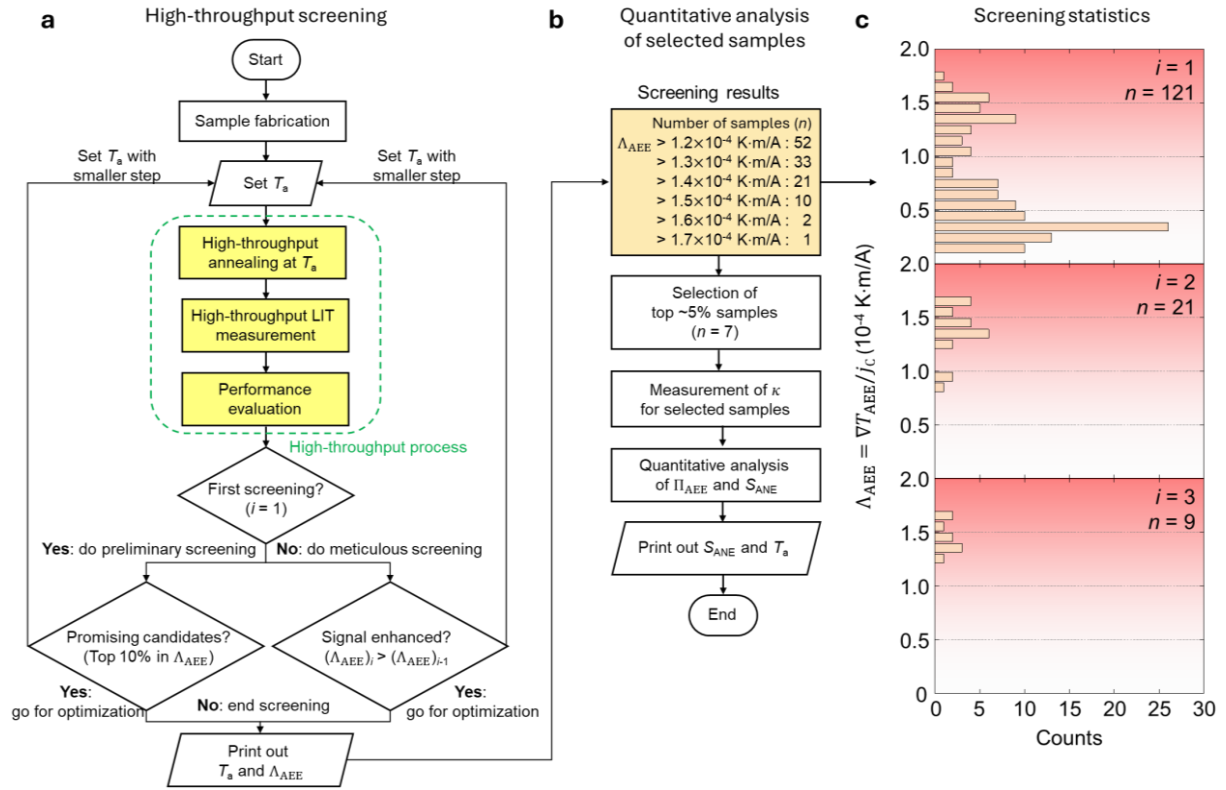

**Fig. 2 | Flowchart of high-throughput screening process for transverse thermoelectric materials and overview of screening results.** (a), (b) Flowcharts for (a) feedback-based high-throughput screening and (b) quantitative analysis of selected samples. (c) Screening statistics obtained from the process.

Next, we detail our approach to the discovery of efficient transverse thermoelectric materials using the high-throughput screening method. We introduce a feedback-based sample development strategy that combines high-throughput screening and quantitative analysis, as outlined in the flow chart shown in **Fig. 2**. Initially, eleven Fe-based amorphous ribbons were prepared using the melt-spinning method (**Methods, Table 1**). These amorphous alloys were selected because their structural changes upon annealing have previously been associated with enhanced ANE performance under carefully controlled $T_a$ [37–39]. Composed primarily of Fe (73–86 at.%) along with approximately 20 at.% of Si and B as glass-forming elements, these alloys are also free of costly rare metals such as Co, Ga, and Sm, making them attractive for scalable and cost-effective thermoelectric applications. The amounts of Nb, Cu, and P were varied systematically to investigate their influence on crystallization behavior and ANE performance. The eleven alloy compositions were categorized into five groups (A–E), as shown in **Table 1**.

Based on the crystallization temperatures determined by differential scanning calorimetry (DSC, data shown in **Fig. S1** and **Table 1**), potential $T_a$ values were selected, ranging from 623 K to 1073 K with 50 K increments, covering the full range necessary for initial screening (iteration $i = 1$). This process yielded 121 samples ($n = 121$) for $i = 1$. Given the large $n$ across different compositions and $T_a$, conventional one-by-one annealing and evaluation methods present significant practical limitations, underscoring the necessity of our high-throughput approach. Sample identifiers were determined based on alloy groups and their respective $T_a$ values (in K). For example, samples labeled A1-300 and D2-673 correspond to the $Fe_{73.5}Si_{13.5}B_9Nb_3Cu_1$ and $Fe_{79}Si_4B_{14}Cu_1P_2$ compositions (Table 1), annealed at 300 K (as-spun) and 673 K, respectively.

The prepared ribbons were then cut into a width of 1–2 mm and a length of approximately 10 mm. Annealing was performed automatically at each selected $T_a$ for 15 min, followed by water quenching using the automatic annealing system (**Fig. 1a** and **Methods**). Samples were covered with tantalum foil during annealing to ensure batch-level thermal uniformity and prevent contamination.

For performance evaluation, samples were placed parallel to each other on a glass substrate within the FOV and electrically connected in series (**Figs. 1b** and **3**). Typically, 7–10 samples could be measured simultaneously. The ANE performance of multiple samples was evaluated under $\mu_0H$ of ± 0.3 T with $\mu_0$ being vacuum permeability, which was sufficiently large to achieve in-plane saturation magnetization. Transverse thermoelectric performance ($\Lambda_{AEE} = \nabla T_{AEE}/j_c$) of 121 samples at $i = 1$ is presented in **Fig. 2c**. Several as-spun and annealed samples at mild $T_a$ of 623–673 K showed notably enhanced performance. From these initial 121 samples, the top 10% performers ($n = 12$) were selected for further optimization in subsequent iterations. The selected samples were A2-300, A2-673, B1-300, B1-623, B1-673, C1-300, C1-623, C3-300, C4-673, D1-623, D1-673, and D2-673. These promising samples were mostly either as-spun or annealed within the $T_a$ range of 623–673 K, consistent with previous studies [37,38].

Further optimization was conducted by repeating the annealing steps with progressively reduced temperature steps ($\Delta T_a$), halving each previous step size. Specifically, the $\Delta T_a$ values at $i = 2$ and $i = 3$ were reduced to 25 K and 12.5 K, respectively. The iterative optimization continued until additional steps did not produce significant improvements in performance (i.e., proceed to the next iteration when $(\Lambda_{AEE})_i > (\Lambda_{AEE})_{i-1}$; otherwise, stop). Each iteration yielded $n$ of 121 ($i = 1$), 21 ($i = 2$), and 9 ($i = 3$), resulting in a total of $n = 151$, demonstrating the efficiency of our screening strategy. Overall screening results at each iteration are summarized in **Fig. 2c**.

Following the high-throughput screening, quantitative analyses were conducted on selected samples exhibiting top ~5% performance (**Fig. 2b**). The transport properties, including $\kappa$, were thoroughly evaluated to accurately quantify the $\Pi_{AEE}$ and $S_{ANE}$ values. Detailed analyses of these promising samples are discussed in **Section "Quantitative evaluation of transverse thermoelectric performance in selected samples"**.

**Table. 1** | Material properties of the samples used in this study. Blanks (-) indicate not relevant properties.

| Sample ID | Composition (at.%) | | | | | | Crystallization temperature (K) | | Specific heat (J/gK) | Density (kg/m³) | Thickness, $t$ (μm) |
|---|---|---|---|---|---|---|---|---|---|---|---|
| | Fe | Si | B | Nb | Cu | P | $T_{x1}$ | $T_{x2}$ | | | |
| A1 | 73.5 | 13.5 | 9 | 3 | 1 | - | 803 | 973 | 0.529 | 7.30 | 23.1±1.4 |
| A2 | 79 | 4 | 14 | 2 | 1 | - | 726 | 855 | 0.536 | 7.39 | 19.9±1.7 |
| A3 | 78 | 4 | 14 | 3 | 1 | - | 735 | 868 | 0.534 | 7.41 | 15.9±1.8 |
| B1 | 82 | 4 | 14 | - | - | - | 748 | 800 | 0.540 | 7.32 | 15.6±2.1 |
| C1 | 81 | 3 | 13 | - | 1 | - | 701 | 796 | 0.423 | 7.35 | 21.3±1.6 |
| C2 | 81 | 4 | 14 | - | 1 | - | 702 | 793 | 0.433 | 7.33 | 21.5±1.9 |
| C3 | 81 | 3 | 15 | - | 1 | - | 703 | 786 | 0.543 | 7.31 | 19.8±1.5 |
| C4 | 81.4 | 4 | 14 | - | 0.6 | - | 708 | 795 | 0.540 | 7.33 | 19.2±1.2 |
| D1 | 80 | 4 | 14 | - | 1 | 1 | 716 | 809 | 0.543 | 7.30 | 10.4±1.8 |
| D2 | 79 | 4 | 14 | - | 1 | 2 | 729 | 817 | 0.546 | 7.27 | 17.4±2.7 |
| E1 | 86 | - | 14 | - | - | - | 667 | 753 | 0.530 | 7.45 | 19.4±1.6 |

**High-throughput evaluation of transverse thermoelectric performance utilizing lock-in thermography**

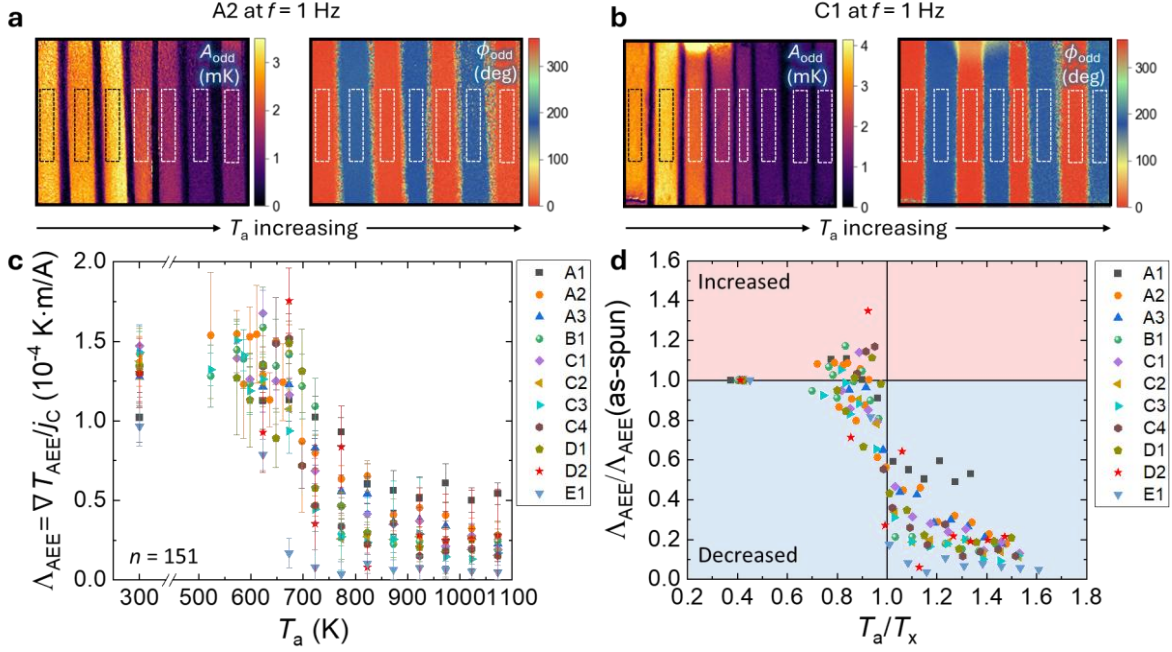

**Fig. 3 | Evaluation of transverse thermoelectric performance utilizing lock-in thermography.** (a), (b) Lock-in amplitude ($A_{odd}$) and phase ($\phi_{odd}$) images for the A2 (a) and C1 (b) samples at a frequency $f$ of 1 Hz, illustrating examples of high-throughput evaluation. The annealing temperatures ($T_a$) increase from left to right in (a) and (b). The leftmost samples are as-spun samples without annealing, with $T_a$ increasing from $T_a$= 350 K in 50 K increments at the first iteration. (c) $\nabla T_{AEE}/j_C$ at $T$=300 K as a function of $T_a$. The data for $T_a$=300 K in (c) correspond to the as-spun samples. (d) Normalized $\Lambda_{AEE} = \nabla T_{AEE}/j_C$ by those of as-spun samples as a function of $T_a/T_{x1}$. The error bars in (c) represent the standard deviation of the measurements and are not shown in (d) to clearly display the increasing trend of transverse thermoelectric signal near $T_a \sim T_x$.

Now, we present the observed high-throughput LIT signals obtained from multiple samples. As described in the previous sections, the samples were placed in parallel on glass substrates and electrically connected in series. The samples shown in **Figs. 3a and 3b** were annealed at varying $T_a$, with the leftmost samples being as-spun (i.e., without annealing), and subsequent samples annealed at progressively higher $T_a$, starting from 350 K in 50 K increments. Due to their parallel configuration, the direction of current flow alternates between adjacent samples. Considering the orthogonal relationship between current and magnetization in the AEE, this configuration leads to alternating directions of $\nabla T_{\mathrm{AEE}}$ across adjacent samples, as directly reflected by the $\phi_{\mathrm{odd}}$ data. The $\phi_{\mathrm{odd}}$ images in **Figs. 3a** and **3b** clearly exhibit ~180° differences between adjacent samples, confirming that the observed $A_{\mathrm{odd}}$ signals originate from AEE without any sign reversal in material response across $T_{\mathrm{a}}$ [49,50,52,53]. Additionally, the constant $\phi_{\mathrm{odd}}$ values of 0 or 180° with minimal deviation (< 3°) suggest well-aligned in-plane magnetization within the FOV under the applied $\mu_0 H$ of 0.3 T. The $A_{\mathrm{odd}}$ signals were extracted by averaging values from the central regions of the samples with flat surfaces (boxed in **Figs. 3a** and **3b**) to minimize defocusing-related errors. By analyzing these $A_{\mathrm{odd}}$ signals from the LIT images, multiple samples could be evaluated simultaneously. For high-throughput estimation of $\Pi_{\mathrm{AEE}}$, all data were analyzed at $f = 1$ Hz, closely approximating steady-state conditions, as confirmed by previous studies [37,39].

**Figure 3c** shows $\Lambda_{\mathrm{AEE}} = \nabla T_{\mathrm{AEE}}/j_{\mathrm{C}}$ for all screened samples, plotted as a function of $T_{\mathrm{a}}$. The data exhibit a consistent trend: the samples annealed near 623–673 K show the highest values, clearly outperforming both their as-spun and fully crystalline counterparts. Notably, the highest $\Lambda_{\mathrm{AEE}}$ value ($1.76 \times 10^{-4}$ K·m/A) was observed in the sample D2-673. The most promising samples identified include A2-523, A2-573, A2-598, A2-611, B1-623, C1-623, and D2-673.

Using these statistical data, we further explore the origins of the enhanced signals in the annealed amorphous alloys. To clarify this trend, **Fig. 3c** was replotted in **Fig. 3d**, showing the normalized $\Lambda_{\mathrm{AEE}}$ values with respect to their as-spun counterparts $\Lambda_{\mathrm{AEE}}$(as-spun), plotted against $T_{\mathrm{a}}/T_{\mathrm{x1}}$. **Figure 3d** shows that most alloys exhibited enhanced $\Lambda_{\mathrm{AEE}}$ when annealed just below $T_{x1}$. Notably, this enhancement occurred consistently across all sample groups, including those with Nb, Cu, or P additions (groups A, C, and D) and even those without (group B).

The analysis of the extensive dataset offers key insights into the origin of transverse electron deflection in these complex systems. First, the enhancement of $\Lambda_{AEE}$ observed in groups A, C, and D, particularly those containing Cu, suggests a possible contribution from interfacial spin-orbit interactions between Cu clusters and the Fe-based amorphous matrix [37,55]. To investigate the structural change upon annealing, we measured the atomic distributions of D2-673 using APT (**Methods**), which exhibited the highest enhancement in $\Lambda_{AEE}$, and D2-300 used as a reference. The APT results in **Fig. S2** reveal a high number density of Cu nanoclusters in an amorphous matrix with an average cluster size of approximately 5 nm in D2-673, while D2-300, the fully amorphous counterpart, shows a uniform atomic distribution without evidence of phase separation. It is noteworthy that this Cu cluster density in our samples ($1.2 \times 10^{24}$ m$^{-3}$) significantly exceeds that reported in previous state-of-the-art flexible ANE materials [37] ($7.3 \times 10^{23}$ m$^{-3}$) with $S_{ANE}$ of 3.7 μV/K, potentially leading to stronger transverse deflections. Second, according to a recently proposed study, electron deflection in disordered systems can also arise from heterogeneous composite formation, causing electron paths to become meandering [38]. Specifically, when two phases with significantly contrasting transport properties are physically mixed, transverse deflection can emerge at their heterointerfaces. The condition for such a mechanism is that one phase should have a lower longitudinal but higher transverse conductivity compared to the other. In the samples annealed near $T_{x1}$, the emergence of multiple domains, including amorphous and crystalline phases, is expected. These domains may include Fe-based amorphous matrix, Fe-based crystalline phases and Cu nanoclusters (**Fig. S2**). Such heterostructures satisfy the criteria for enhanced ANE, as the long-range ordered phases (Fe alloys and Cu clusters) exhibit higher longitudinal and lower transverse conductivity compared to the disordered amorphous matrix, where electrons are localized by scattering with short-range phonons [56–58]. This explanation is further supported by the thermal conductivity trends discussed in the next section. Lastly, the results from the subtle differences in composition across the sample groups highlight the importance of the composition and local atomic order of the amorphous matrix in generating transverse deflection. Notably, the enhancement in $\Lambda_{AEE}$ was observed even in Cu-free samples (group B), indicating that Cu clustering is not the sole contributor to the enhanced $\Lambda_{AEE}$. The APT results obtained from B1-623 in **Fig. S3** show no clear evidence of phase separation, and thus the nanostructural heterogeneity alone cannot fully explain this enhancement as well. The STEM results for B1-623 and D2-673 in **Fig. S4** exhibit signatures of possible short-range ordering within the disordered matrix (**Methods**). This observation suggests that subtle compositional tuning and/or short-range ordering within the amorphous matrix may also play an important role in transverse deflection, which requires further theoretical investigation considering the statistical thermodynamics in such disordered solids, where experimental detection of these changes is inherently challenging.

**Quantitative evaluation of transverse thermoelectric performance in selected samples**

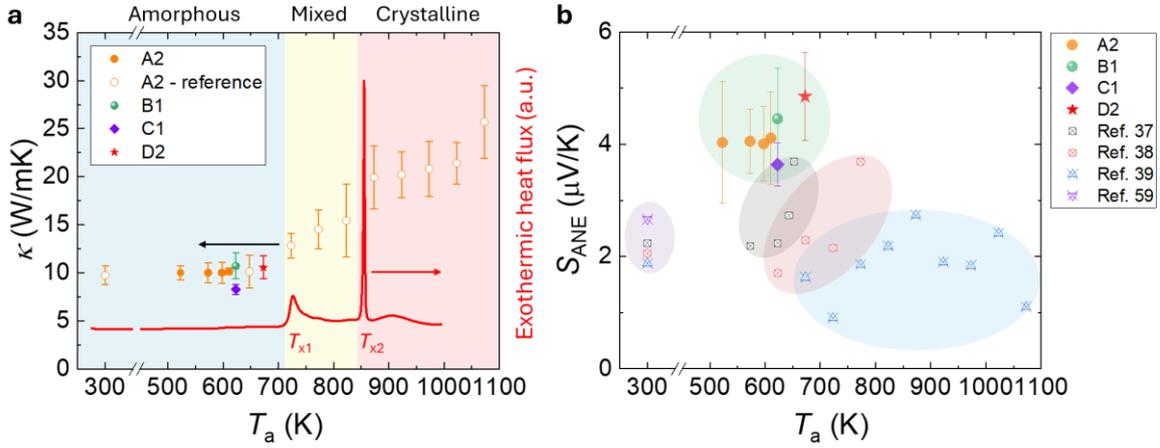

**Fig. 4 | Quantitative evaluation of transport properties in selected samples.** (a) Thermal conductivity ($\kappa$) and (b) anomalous Nernst coefficient ($S_{ANE}$) as a function of $T_a$.

Now we provide a detailed quantitative analysis for the selected samples (**Fig. 2b**). We evaluated $\kappa$ of selected samples by multiplying the thermal diffusivity, specific heat, and density (**Fig. S5** and **Table 1**). The thermal diffusivity was measured by a laser-based LIT technique (**Methods**). All samples exhibited isotropic thermal diffusivity due to their amorphous or polycrystalline nature (**Fig. S5**). **Figure 4a** shows $\kappa$ for the selected samples, alongside data for reference samples from the A2 group annealed at various $T_a$. For the reference A2 group, $\kappa$ increases in two distinct temperature ranges: 678 K < $T$ < 723 K, and 823 K < $T$ < 873 K. These ranges align well with the crystallization temperatures at $T_{x1}$ = 726 K and $T_{x2}$ = 855 K (**Fig. 4a**). Specifically, $\kappa$ is ~10 W/mK below $T_{x1}$ (amorphous regime), ~14 W/mK between $T_{x1}$ and $T_{x2}$ (amorphous-crystalline mixed regime), and ~22 W/mK above $T_{x2}$ (fully crystallized regime). These observed trends in $\kappa$ are consistent with expectations that crystallization and grain growth enhance both electronic and phononic contributions to thermal transport by reducing scattering from structural disorder. All the selected samples exhibit $\kappa$ values of 8–9 W/mK, indicating that they remain in the amorphous regime regardless of composition.

The $\Pi_{AEE}$ and $S_{ANE}$ values of the selected seven samples were quantitatively calculated using Eqs. (1) and (3). The estimated $\Pi_{AEE}$ values range from 1.09 to 1.45 mV, corresponding to the $S_{ANE}$ values of 3.6–4.8 µV/K via Onsager reciprocity (**Fig. 4b**). Notably, the D2-673 sample exhibited the maximum value of 4.8 µV/K, associated with a large density of Cu clusters formed in amorphous matrix (**Fig. S2**). These values are the highest reported among flexible materials in both bulk and thin-film forms, demonstrating the efficient development of large transverse thermoelectric materials through feedback-based high-throughput screening. **Figure 4b** shows the comparison of performance with flexible bulk materials from the literature, including flexible Co-based amorphous metals [59], Fe-based amorphous metals, with [37] and without [38] Cu, and Fe-Pt-based amorphous metals [39]. We note that bulk materials are advantageous for energy harvesting compared to thin films, whose output power is limited by their inherently large resistance. The detailed comparisons including both flexible bulk and thin-film forms are provided in **Fig. S6**, including an Fe-Ga thermopile on a polyethylene terephthalate substrate [60], Sm-Co-based amorphous alloy on a polyethylene naphthalate substrate [61], Fe-Al thermopile on a polyimide substrate [54]. These comparisons confirm that the selected materials exhibit record-high ANE performance among flexible materials, highlighting the effectiveness of our high-throughput approach in identifying top-performing candidates.

This structural transition impacts not only thermal transport properties, but also mechanical flexibility. Typically, amorphous alloys offer superior mechanical flexibility due to the ductility of the amorphous phase. However, it is well-known that increasing $T_a$ enhances crystallinity and consequently increases brittleness, compromising flexibility. To assess the mechanical flexibility of selected samples annealed at mild $T_a$, we attached the samples to a curved surface with radius of 10 mm and confirmed the flexibility in the high-performance samples annealed at $T_a < T_{x1}$ (**Fig. S7**), whereas the samples annealed at higher $T_a$ lost their flexibility due to increased crystallinity. The flexibility of the high-performance samples suggests their applicability for energy harvesting from curved or irregularly shaped heat sources, enabling advanced designs of thermoelectric devices such as rolled-up structures proposed in [16].

**Conclusion**

In summary, we have demonstrated a high-throughput material screening strategy for discovering efficient flexible transverse thermoelectric materials by integrating an automated annealing system with LIT. A total of 151 amorphous alloy ribbons, with varying compositions and annealing conditions, were fabricated and evaluated through the feedback-based iterative screening. From this large dataset, seven promising candidates were identified, exhibiting large ANE performance reaching 4.8 µV/K, the highest reported among flexible amorphous materials. These high-performance samples maintained mechanical flexibility, making them promising for applications involving curved or irregularly shaped heat sources.

A statistical evaluation of the full dataset revealed a universal enhancement in transverse thermoelectric conversion near the first crystallization temperature, regardless of composition. The largest enhancement was observed in the $Fe_{79}Si_4B_{14}Cu_1P_2$ sample annealed at 673 K (D2-673), where a large density of Cu nanoclusters was precipitated within the Fe-based amorphous matrix as confirmed by APT. Notably, the enhancement was observed even in the Cu-free $Fe_{82}Si_4B_{14}$ sample annealed at 623 K (B1-623), despite the absence of Cu clusters or Fe-based nanocrystals. This underscores that subtle compositional tuning within the amorphous matrix and local atomic order also play a crucial role in transverse thermoelectric conversion. This points to the need for further theoretical development to understand transverse deflection in disordered systems, where conventional models based on Berry curvature are insufficient.

Our high-throughput screening methodology, coupled with robust statistical analysis and detailed structural characterization, provides a comprehensive platform for advancing the development of next-generation flexible ANE materials. Moreover, the resulting dataset lays the groundwork for data-driven materials discovery using machine learning, especially in complex amorphous systems where experimental intuition and first-principles modeling alone face limitations.

**Methods**

*Preparation of amorphous ribbons*

The master alloy ingots were prepared by melting a mixture of high-purity elements using a high-frequency induction-melting furnace under an Ar atmosphere. Subsequently, the ingot was crushed into small pieces and charged into a quartz tube with a 5.0 mm × 0.4 mm nozzle opening. The amorphous ribbons were then produced using the single-roll melt-spinning technique. This process involved melting the ingot pieces using induction heating and ejecting the molten from the nozzle under an Ar atmosphere at a pressure of 0.02 MPa onto a rotating Cu wheel. The wheel speed was optimized based on the alloy composition, ranging from 30 to 35 m/s, to obtain high-quality amorphous ribbons. The gap between the nozzle and Cu wheel was maintained at 0.2 mm.

*Sample annealing using automatic robotic annealing system*

The amorphous samples were annealed using a customized automatic robotic system (YSD mechatro systems) at various temperatures under a high-vacuum condition (~ $10^{-2}$ Pa). The system conducts the heat treatment of the samples by the following sequence. First, the robotic arm grabs the samples loaded in a quartz tube and vacuum inside through the vacuum line connected to a high vacuum pump. The robotic arms and quartz tube are connected through a rubber O-ring for vacuuming. During the vacuuming, the furnace increases the temperature at $T_a$. When the vacuum level becomes lower than a specific value (set as 0.02 Pa) and the temperature of the furnace becomes stable, the samples are transferred into the furnace and hold for 15 min. Then, the sample is transferred to a quenching bath and water quenched.

During the water quenching, the atmosphere inside was partially filled by inert gas Ar to maximize the quenching efficiency by convection. The water cooling was conducted for 15 min and then the sample was transferred to the initial position on the sample holder and flushed by air for sample exchange. All the processes, including the sample transfer, evacuating, and quenching, are automatically conducted by a centralized unit with an automatic program. The samples were covered with a thin tantalum foil to prevent potential contamination inside the tube. The total 3 units of quenching systems were used for the high-throughput sample annealing, which can load 12 independent samples with different annealing conditions in total with 3 sample transfer systems, 3 furnaces and 3 quenching baths, as illustrated in **Fig. 1**. The temperature of the furnaces was controlled by a thermocouple inside of each furnace and the annealing temperatures of the samples were estimated by pre-calibrated temperature data using thermocouples inside the quartz tube.

*Analysis of dynamic structure change during the heating*

The exothermic reaction due to structural change was investigated using DSC (Rigaku, Thermo plus EVO2). The DSC measurement was conducted with a Pt pan at temperatures ranging from 273 K to 993 K with a ramp rate of 10 K/min. The samples were baked at 373 K for 10 min before the measurement to remove moisture.

*Characterization of transport properties using lock-in thermography*

AEE of all samples was measured using the LIT system (DCG Systems Inc., ELITE). The samples were fixed on a glass substrate using an insulating adhesive and connected electrically in series using silver epoxy. The measurements were conducted by applying a square-wave-modulated charge current with an amplitude of 30-50 mA to the samples of various thickness under an in-plane magnetic field of ±0.3 T at room temperature ($T$ = 300 K). Sample thickness ranged from 10.4 to 23.1 μm (**Table 1**), measured using a micrometer.

The thermal diffusivity $D$ was also assessed using the LIT method [37]. The setup includes an infrared camera, diode laser, function generator, and lock-in analysis system. The diode laser, modulated by a reference signal at $f$ from the function generator, directs a focused spot laser beam onto the backside of the sample. This laser spot generates heat waves that spread radially within the sample. As the heat waves propagate, the IR camera captures the thermal response on the front side of the sample. The LIT system processes these thermal images to provide the spatial distribution of the $A$ and $\phi$ signals synchronized with $f$. By examining the relationship between $\phi$ and the distance $r$ from the laser point heat source, $D$ can be determined. The formula used for calculating $D$ is

$$D = \frac{\pi f}{\left(\frac{d\phi}{dr}\right)^2}.$$

For the experiment, a high precision LIT system (DCG Systems Inc., ELITE) was used along with a diode laser (Omicron Inc., LDM637D.300.500) with a wavelength of 638 ± 1 nm. The laser beam was focused using an optical setup, resulting in an approximately 7-μm spot in a diameter, operating at a power of 20 mW. The beam was modulated at $f$ = 7.5 Hz, which was selected to optimize the signal-to-noise ratio and minimize heat losses, ensuring that the thermal diffusion length (= $D/\pi f$) remained away from the edges of the sample [62].

***Structural analysis via APT and TEM***

Elemental distribution was observed using APT in laser mode with a CAMCECA LEAP 5000 XS instrument, operated at a base temperature of 30 K and a laser pulse rate and energy of 250 KHz and 30 pJ, respectively. APT data analysis was performed using CAMECA AP Suit 6.1 software. Microstructural analysis was conducted using a FEI Titan G2 80-200 TEM. APT and TEM specimens were prepared using the lift-out technique with a FEI Helios 5UX dual beam-focused ion beam.

**Author Contributions**

S.J.P. designed and conceived the study. K.U. supervised the project. R.G. and H.S.A. fabricated the initial as-spun amorphous metals, performed APT and STEM, and analyzed the data. S.J.P. conducted the high-throughput heat treatment of the samples using the automatic robotic annealing system, measured and analyzed the DSC data, and performed the high-throughput performance evaluation by measuring the LIT and analyzing the data. A.A. and H.N. measured the thermal diffusivity of the samples using LIT. F.A. and T.H. supported the transport measurements. S.J.P. wrote the manuscript with input from all authors.

**CRediT authorship contribution statement**

**Sang Jun Park** (Conceptualization: Lead; Data curation: Lead; Formal analysis: Lead; Investigation: Lead; Methodology: Lead; Validation: Lead; Visualization: Lead; Writing – original draft: Lead; Writing – review & editing: Lead)

**Ravi Gautam** (Investigation: Supporting; Visualization: Supporting; Writing – review & editing: Supporting)

**Abdulkareem Alasli** (Investigation: Supporting; Writing – review & editing: Supporting)

**Takamasa Hirai** (Investigation: Supporting; Writing – review & editing: Supporting)

**Fuyuki Ando** (Investigation: Supporting; Writing – review & editing: Supporting)

**Hosei Nagano** (Investigation: Supporting; Writing – review & editing: Supporting)

**Hossein Sepehri-Amin** (Investigation: Supporting; Writing – review & editing:

Supporting)

**Ken-ichi Uchida** (Investigation: Equal; Methodology: Equal; Project administration: Lead; Resources: Lead; Supervision: Lead; Writing – review & editing: Supporting)

**Declaration of competing interest**

The authors declare that they have no known competing financial interests or personal relationships that could have appeared to influence the work reported in this paper.

**Acknowledgements**

The authors thank Tadakatsu Ohkubo for cooperation of using the automatic robotic annealing system and Yoshiya Toyooka, Hiroyuki Sebata and Mizue Isomura for technical supports. This work was supported by ERATO "Magnetic Thermal Management Materials" (grant no. JPMJER2201) from JST, Japan.

**Appendix A. Supplementary data**

**Data availability**

Data will be made available on request.

# Supplementary data for

# High-throughput development of flexible amorphous materials showing large anomalous Nernst effect via automatic annealing and thermoelectric imaging

Sang J. Park[1,*], Ravi Gautam[1], Abdulkareem Alasli[2], Takamasa Hirai[1], Fuyuki Ando[1], Hosei Nagano[2], Hossein Sepehri-Amin[1] and Ken-ichi Uchida[1,3,*]

[1] National Institute for Materials Science, Tsukuba 305-0047, Japan
[2] Department of Mechanical Systems Engineering, Nagoya University, Nagoya 464-8601, Japan
[3] Department of Advanced Materials Science, Graduate School of Frontier Sciences, The University of Tokyo, Kashiwa 277-8561, Japan

*Corresponding authors: PARK.SangJun@nims.go.jp (S.J.P.); UCHIDA.Kenichi@nims.go.jp (K.U.)

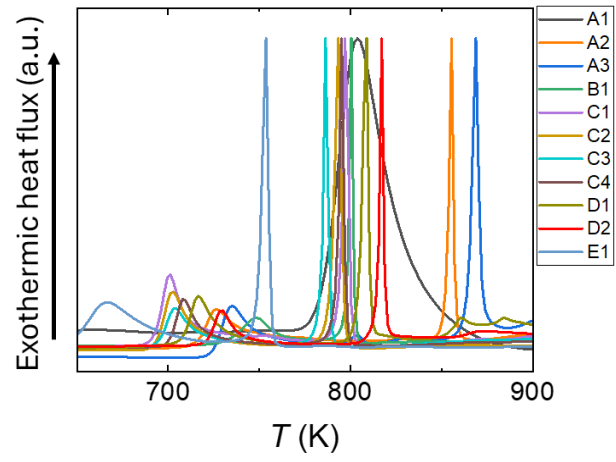

**Fig. S1 | Crystallization of the samples, investigated by differential scanning calorimetry.**

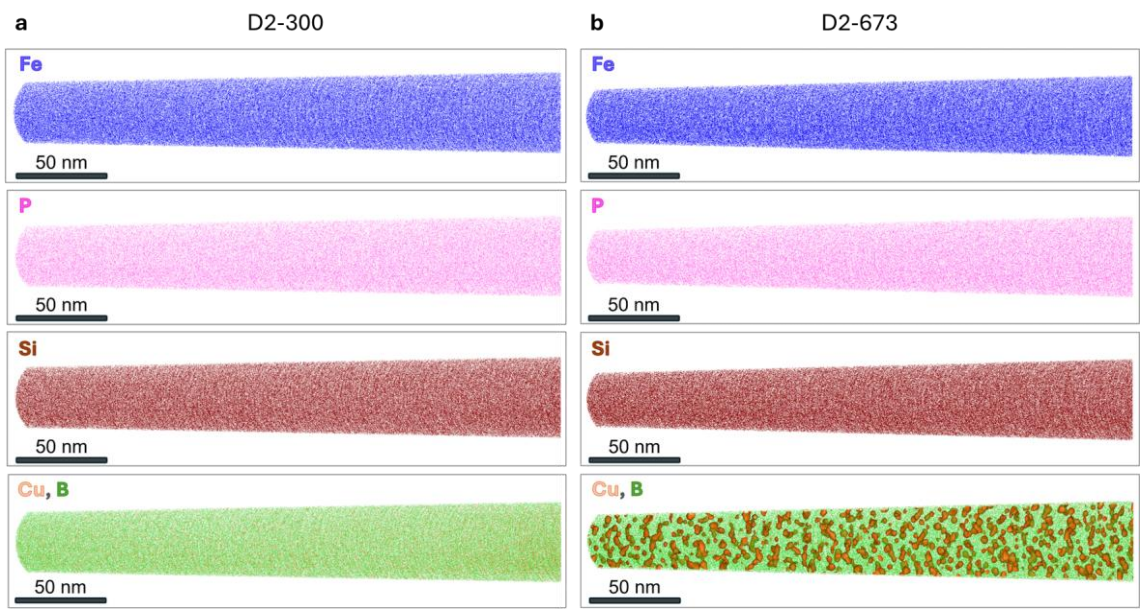

**Fig. S2 | Atomic probe topography analysis of (a) D2-300 and (b) D2-673 samples.** APT maps of the constituent elements, Fe (blue), P (pink), Si (dark red), B (green), and Cu (orange), shown in 10 nm sliced box from whole data for a better visualization.

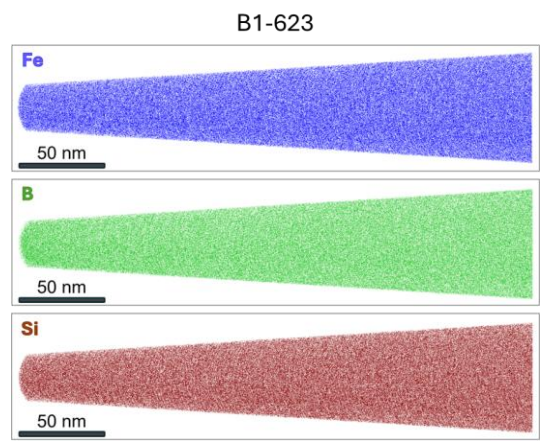

**Fig. S3 | Atomic probe topography analysis of B1-623 samples.** APT maps of the constituent elements, Fe (blue), Si (dark red), B (green), shown in 10 nm sliced box from whole data for a better visualization.

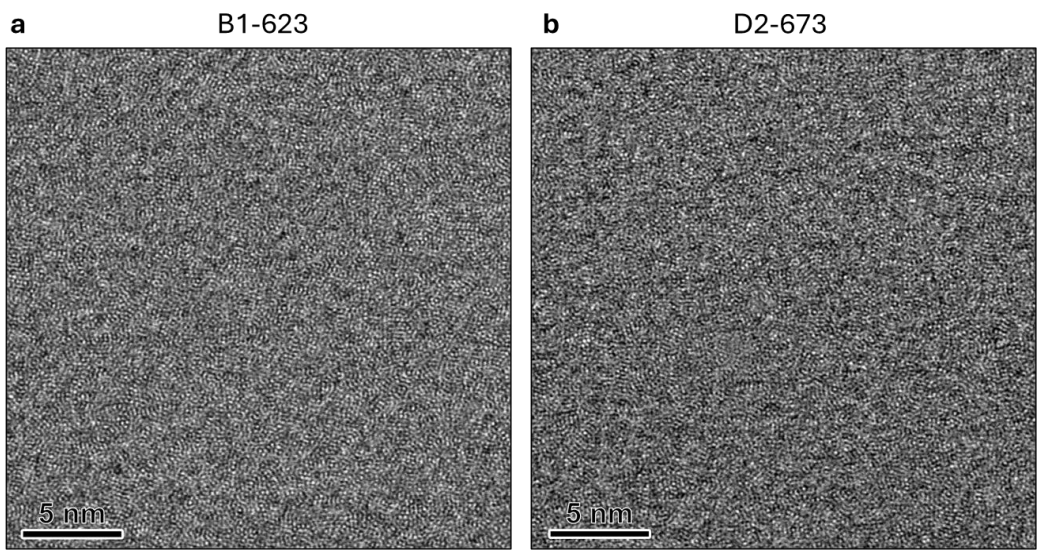

**Fig. S4 | Scanning transmission electron microscopy images of (a) B1-623 and (b) D2-673 samples.**

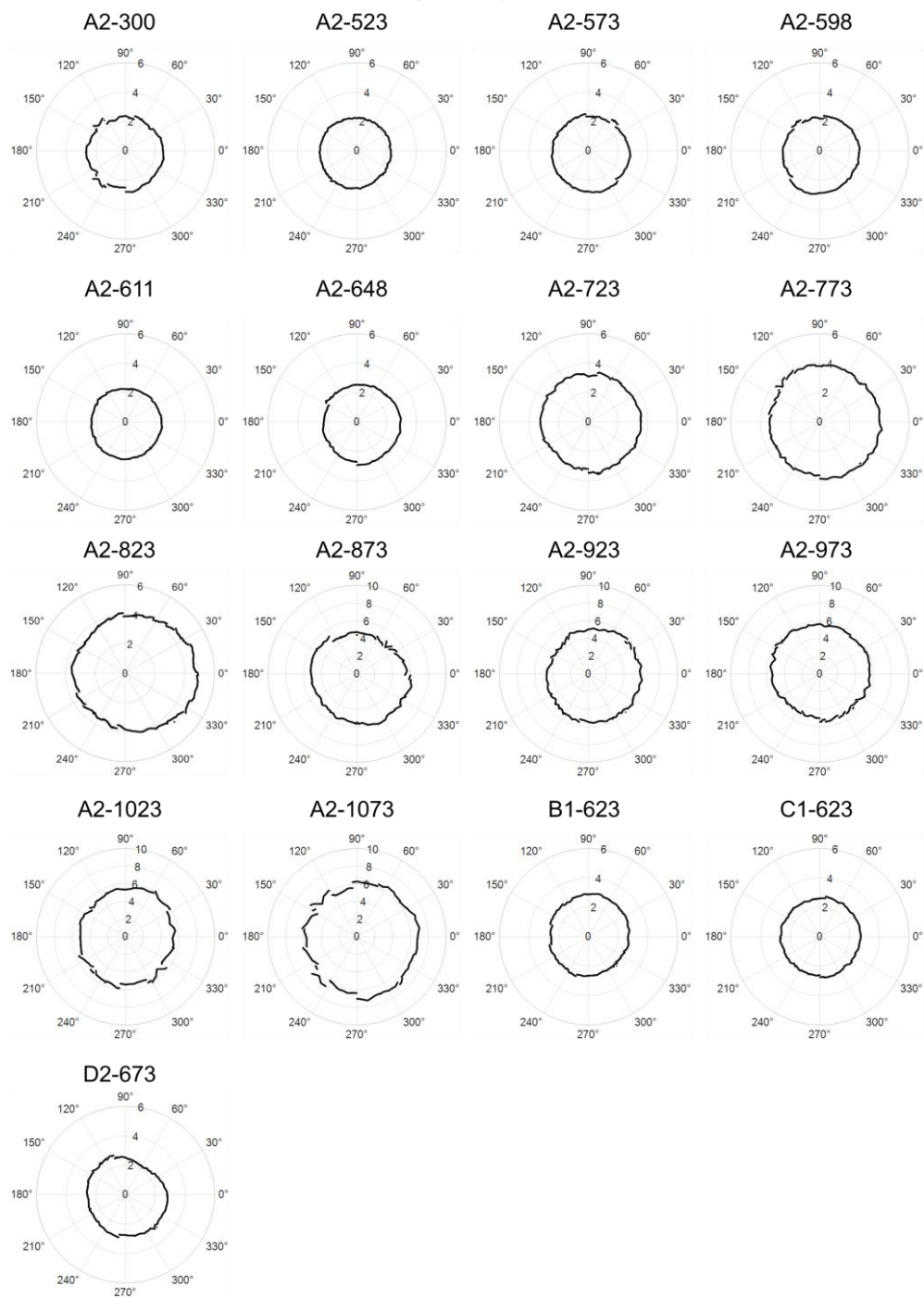

**Fig. S5 | Spatial distribution of thermal diffusivity of the samples presented in Fig. 4a in the main text.**

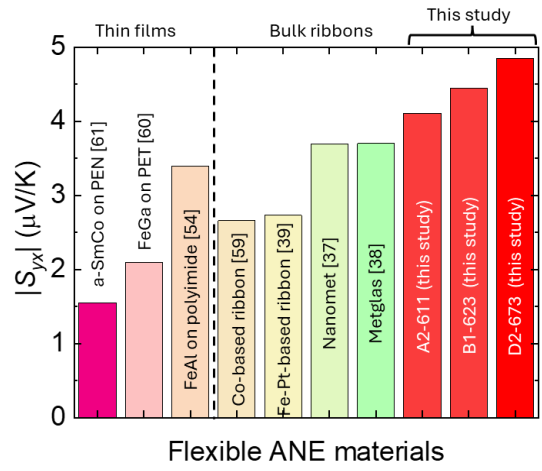

**Fig. S6 | Comparison of flexible ANE materials, including thin films and bulk ribbons.** The reference numbers correspond to those in the main text.

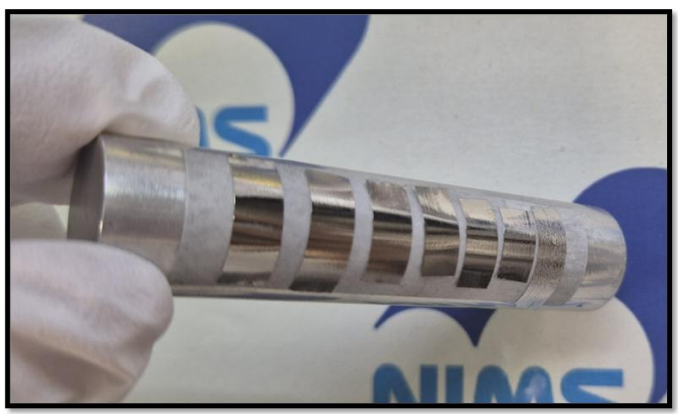

**Fig. S7 | Flexibility test.** The samples are A2-523, A2-573, A2-598, A2-611, B1-623, C1-623, and D2-673 from left to right and attached on a 10-mm radius curve.